\documentstyle[prl,aps,psfig]{revtex}
\newcommand{\beq}{\begin{equation}}
\newcommand{\eeq}{\end{equation}}
\newcommand{\beqa}{\begin{eqnarray}}
\newcommand{\eeqa}{\end{eqnarray}}
\newcommand{\ba}{\begin{array}}
\newcommand{\ea}{\end{array}}

\begin{document}
\draft

\twocolumn[\hsize\textwidth\columnwidth\hsize\csname
@twocolumnfalse\endcsname

\widetext 
\title{Bose Condensate in a Double-Well Trap: \\
Ground State and Elementary Excitations} 
\author{L. Salasnich$^{(*)(+)}$, A. Parola$^{(*)(++)}$ 
and L. Reatto$^{(*)(+)}$} 
\address{$^{(*)}$ Istituto Nazionale per la Fisica della Materia, 
Unit\`a di Milano, \\
Via Celoria 16, 20133 Milano, Italy \\
$^{(+)}$ Dipartimento di Fisica, Universit\`a di Milano, \\
Via Celoria 16, 20133 Milano, Italy \\ 
$^{(++)}$ Dipartimento di Scienze Fisiche, Universit\`a dell'Insubria, \\ 
Via Lucini 3, 23100 Como, Italy}
\maketitle
\begin{abstract}
We study the Bose-Einstein condensate in the MIT double-well trap. 
We calculate the ground state density profile of $^{23}$Na atoms 
and the Bogoliubov spectrum of the elementary excitations 
as function of the strength of the double-well barrier. 
In particular, we analyze the behavior of 
quantum-mechanical collective excitations. Finally, we discuss the 
observability criteria for macroscopic quantum tunneling and macroscopic 
quantum self-trapping. 
\end{abstract}
\pacs{03.75.Fi, 05.30.Jp, 32.80.Pj}

]

\narrowtext

\par
Three years ago the Bose-Einstein condensation of 
alkali vapors $^{87}$Rb, $^{23}$Na and $^{7}$Li has been achieved 
in magnetic harmonic traps at temperature of the order of $100$ nK [1-3]. 
Theoretical studies of the Bose-Einstein condensate (BEC) in 
harmonic traps have been performed for the ground state [4-8], 
collective low-energy surface excitations [9-11] and vortex states [6,12]. 
\par 
In this paper we study the BEC in the MIT double-well trap 
given by a harmonic anisotropic potential plus a Gaussian 
barrier along the $z$ axis, which models the effect of a laser 
beam perpendicular to the long axis of the condensate. 
In the MIT experiment [13], the macroscopic interference 
of two Bose condensates released from the double minimum 
potential has been demonstrated. Such phenomenon has been theoretically 
reproduced [14] by using the Gross-Pitaevskii (GP) equation [15]. 
Here, we concentrate on the ground state properties of 
the condensate and calculate the spectrum of 
the Bogoliubov elementary excitations as a function of 
the intensity of the laser field. A comparison between our 
calculations and future experiments will clarify the accuracy of the GP
equation and the role of correlations in Bose condensates with up
to $4\times 10^6$ particles.
\par
By varying the strength of the barrier one can observe 
macroscopic quantum effects, like the formation of two Bose condensates, 
the collective oscillations and the quantum tunneling. 
\par 
The Gross-Pitaevskii energy functional [15] of the BEC reads: 
\beq
{E\over N} = \int d^3{\bf r} \;  
{\hbar^2\over 2m} |\nabla \Psi ({\bf r})|^2 
+ V_{ext}({\bf r}) |\Psi ({\bf r})|^2 +{g N\over 2} |\Psi ({\bf r})|^4 \; ,
\eeq 
where $\Psi ({\bf r})$ is the wave function of the condensate
normalized to unity, $V_{ext}({\bf r})$ is the external potential 
of the trap, and the interatomic potential 
is represented by a local pseudopotential so that $g={4\pi \hbar^2 a_s/m}$ 
is the scattering amplitude ($a_s$ is the s-wave scattering length). 
$N$ is the number of bosons of the condensate and $m$ is the atomic 
mass. The extremum condition for the energy functional 
gives the GP equation 
\beq
\Big[ -{\hbar^2\over 2m} \nabla^2 
+ V_{ext}({\bf r}) + g N |\Psi ({\bf r})|^2 \Big] \Psi ({\bf r}) 
= \mu \Psi ({\bf r}) \; ,
\eeq
where $\mu$ is the chemical potential. This equation has the form of 
a nonlinear stationary Schr\"odinger equation. 
\par
We study the BEC in an external potential 
with cylindrical symmetry, which is given by 
\beq
V_{ext}({\bf r})={m\omega_{\rho}^2\over 2}\rho^2+{m\omega_z^2 \over 2}z^2 
+ U_0 \exp{ \Big( {-z^2\over 2 \sigma^2} \Big)} \; , 
\eeq 
where $\rho=\sqrt{x^2+y^2}$, $z$ and the angle $\theta$ 
are the cylindrical coordinates. 
The parameter values appropriate for Ref. [13] are 
$\omega_{\rho}=2\pi \times 250$ Hz, $\omega_z=2\pi \times 19$ Hz, 
and $\sigma= 6$ $\mu m$. The anisotropic harmonic trap 
implies a cigar-shaped condensate ($\lambda = \omega_z/\omega_{\rho}=
15/250 <1$), where $z$ is the long axis, and the Gaussian barrier 
of strength $U_0$ creates a double-well potential. 
\par
We perform the numerical minimization of the GP functional 
by using the steepest descent method [18]. 
It consists of projecting onto the minimum of the 
functional an initial trial state by propagating it in imaginary time. 
At each time step the matrix elements entering 
the Hamiltonian are evaluated by means of finite-difference approximants 
using a grid of $200\times 800$ points. 
In our calculations we use the $z$-harmonic oscillator units. 
For $^{23}$Na atoms, the harmonic length is 
$a_z=(\hbar / (m \omega_z))^{1/2} =4.63\;\mu$m 
and the energy is $\hbar \omega_z = 0.78$ peV. 
Moreover, we use the following value for the 
scattering length: $a_s=3$ nm [3]. 
Most of our computations has been performed for $N=5\times 10^{6}$ atoms, 
a value typical of the MIT experiment [13]. 
\par 
In Figure 1 we show the ground state density profile of the $^{23}$Na 
condensate for different values of the strength of the barrier. 
By increasing the strength, the fraction of $^{23}$Na atoms 
decreases in the central region and the Bose condensate separates 
in two condensates. As shown in Table 1, the condensate slightly 
expands in the $z$ direction due to the barrier potential at the origin. 
The numerically calculated density profiles are in good agreement 
with the phase-contrast images of the MIT experiment [13] 
and with the Thomas-Fermi (TF) approximation, 
which neglects the kinetic term in the GP equation. 
Due to the large number of atoms involved ($N=5\times 10^6$), only 
near the borders of the wave function there are small deviations 
from the TF approximation. 
Note that the potential barrier $U_0$ can be written 
as $U_c/k_B=(37 \mu K)P/\sigma^2$ ($\mu$m$^2$/mW), 
where $P$ is the total power of the 
laser beam perpendicular to the long axis of the condensate and 
$\sigma = 6 \mu$m is the beam radius [19]. 
The conversion factor is $0.09$ mW/($\hbar \omega_z$), 
such that $U_0=100$ (in $\hbar \omega_z$ units) 
gives a laser Power $P=9$ mW. 
\par 
Another important property of the BEC is the spectrum 
of elementary excitations. To calculate the energy and wavefunction 
of the elementary excitations, 
one must solve the so-called Bogoliubov-de Gennes (BdG) equations 
[20-22]. The BdG equations can be obtained from the linearized 
time-dependent GP equation. Namely, one can look for $q$ angular momentum
solutions of the form 
$$
\Psi ({\bf r},t) = e^{-{i\over \hbar}\mu t} 
\big[ \psi (\rho ,z) + e^{i q \theta} 
\big( u(\rho ,z) e^{-i \omega t} 
+v^*(\rho ,z) e^{i \omega t} \big) \big] \; 
$$
corresponding to small oscillations of the wavefunction 
around the ground state solution $\psi$. 
By keeping terms linear in the complex functions $u$ and $v$, one finds 
the following BdG equations 
$$
H_{eff}u(\rho,z)
+ g N |\psi (\rho,z)|^2 v(\rho,z) = \hbar \omega \; u(\rho,z) \; ,
$$
\beq 
H_{eff}v(\rho,z)
+ g N |\psi (\rho,z)|^2 u(\rho,z) = -\hbar \omega \; v(\rho,z) \; ,
\eeq
where
$$
H_{eff}=
 -{\hbar^2\over 2m} 
\Big({\partial^2 \over \partial \rho^2} + 
{1\over \rho} {\partial \over \partial \rho} + 
{\partial^2 \over \partial z^2} \Big) 
$$ 
$$
+ {\hbar^2 q^2 \over 2 m \rho^2} + V_{ext}(\rho,z) - \mu 
+ 2g N |\psi (\rho,z)|^2 \; .
$$
The BdG equations allow one to calculate the eigenfrequencies $\omega$ 
and hence the energies $\hbar \omega$ of the elementary 
excitations. This procedure is equivalent to the 
diagonalization of the N-body Hamiltonian of the system in the Bogoliubov 
approximation [22]. The excitations can be classified according to 
their parity with respect to the symmetry $z\to -z$.
\par 
We have solved the two BdG eigenvalue equations by finite-difference 
discretization with a lattice of $40\times 40$ points 
in the $(\rho,z)$ plane. In this way, the eigenvalue problem reduces to the 
diagonalization of a $3200\times 3200$ real matrix. 
We have tested our program in simple models by comparing numerical
results with analytical solutions and 
verified that a $40 \times 40$ mesh already gives reliable
results for the lowest part of the spectrum [12]. 
\par 
In Table 2 we show the $q=0$ lowest elementary excitations of 
the Bogoliubov spectrum for the ground state of the system. 
When the Gaussian barrier is switched off, 
one observes the presence of an odd excitation at energy quite 
close to $\hbar \omega =1$ (in units 
$\hbar \omega_z$). This mode is related to the oscillation of the 
center of mass of the condensate along the z-axis, due to the harmonic 
confinement. This collective oscillation 
is an exact eigenmode of the problem characterized by the 
frequency $\omega_z$, independently of the strength of the interaction. 
The inclusion of the Gaussian barrier modifies the harmonic 
confinement along the $z$-axis 
and this odd collective mode decreases by increasing the strength 
of the barrier (see Figure 2). 
\par
In cylindrical coordinates, 
the other collective mode of the center of mass, due to the 
harmonic confinement along the radial axis, is an 
off-axial oscillation with angular quantum number $q=1$. 
Such oscillation is the lowest even mode for $q=1$ and
in the absence of a gaussian barrier, 
it is exactly equal to the radial frequency 
$\omega_{\rho}$ ($\hbar \omega_{\rho}=250/19=13.158$, 
in $\hbar \omega_z$ units). 
One expects that this mode is only weakly affected by  
the gaussian barrier along the z-axis. 
In Table 3 are reported the first elementary excitations for $q=1$. 
The lowest $q=1$ excitation ($\hbar \omega = 13.132$) remains constant 
and differs by less than $2.5$ $^{o}/_{oo}$ from the theoretical prediction. 
Also when the BEC separates in two condensates, each condensate 
has the same off-axial ($q=1$) collective oscillation 
of the center of mass. 
\par
As shown both in Table 2 and Table 3, 
for large values of the Gaussian barrier, i.e. when the 
BEC separates in two condensates, we find quasi-degenerate pairs 
of elementary excitations (even-odd). 
The lowest $q=0$ mode and the ground-state of the GP equation 
constitute one of such pairs and 
get closer and closer as the barrier is increased. This is not surprising 
because in the infinite barrier limit we have two equal and independent Bose 
condensates with the same energy spectrum. 
\par 
An interesting aspect of BEC in double-well traps is the 
possibility to detect the macroscopic 
quantum tunneling (MQT). The MQT has been recently investigated 
by Smerzi et al. [23,24]. They have found 
that the time-dependent behavior of the condensate in the tunneling 
energy range can be described by the two-mode equations 
\beq
{\dot z}=-\sqrt{1-z^2}\sin{\phi} \; , \;\;\;\;\;\;\; 
{\dot \phi}=\Lambda z +{z\over \sqrt{1-z^2}}\cos{\phi} \; ,
\eeq 
where $z=(N_1-N_2)/N$ is the fractional population 
imbalance of the condensate in the two wells, 
$\phi=\phi_1-\phi_2$ is the relative phase 
(which can be initially zero), and $\Lambda = 4 E^{int}/\Delta E^0$. 
$E^{int}$ is the interaction energy of the condensate 
and $\Delta E^0$ is the kinetic+potential energy 
splitting between the ground state and the quasi-degenerate 
odd first excited state of the GP equation. 
For a fixed $\Lambda$ ($\Lambda >2$), 
there exists a critical $z_c =2\sqrt{\Lambda -1}/\Lambda$ 
such that for $0 < z << z_c$ there are Josephson-like 
oscillations of the condensate with period 
$\tau = \tau_0/\sqrt{1+\Lambda}$, where $\tau_0 =2\pi \hbar /\Delta E^0$. 
But for $z_c<z \leq 1$ there is macroscopic quantum 
self-trapping (MQST) of the condensate: even if the populations
of the two wells are initially set in an asymmetric state ($z\ne 0$)
they maintain the original population imbalance without transferring
particles through the barrier as expected for a free Bose gas.
This two-mode approximation seems quite reliable. In fact, 
we have compared the predictions of the two-mode equations 
with the numerical solutions of the 1D time-dependent GP equation 
in different regimes, finding a very good agreement 
(relative difference in the period of the Josephson-like oscillations 
less than $1$ $^{o}/_{o}$). By using the 1D time-dependent GP equation, 
we have studied the dynamics of the condensate also 
outside the tunneling region, i.e. when the chemical potential 
is higher than the gaussian barrier and the two-mode equations 
do not hold. In such case, starting, 
for example, with the condensate in one well, it 
does not oscillate nor remains self-trapped but instead 
spreads over the two wells. 
\par 
By solving the stationary GP equation 
in the MIT double-well trap with $^{23}$Na, 
we find that the parameter $\Lambda$ is larger than $10^4$ also when few 
particles are present. Nevertheless, we can control 
the dynamics of the condensate by reducing the scattering 
length $a_s$ and the thickness $\sigma$ of the laser beam. 
In Table 4 it is shown that, as expected, 
the parameter $\Lambda$ scales linearly with $a_s$. 
This is an important point because recently it was confirmed 
experimentally the fact that it is now possible to control the 
two-body scattering length by placing atoms in an external 
field [25]. This fact opens the way to a direct observation of
a macroscopic quantum tunneling of thousands of atoms through
a potential barrier.
\par 
Note that also the geometry and the dimensions 
of the trap play an important role. 
In fact, if the condensate is less cigar-shaped 
(greater $\lambda =\omega_z/\omega_{\rho}$) the system 
has a lower chemical potential and weaker gaussian barriers
are requested to enter the tunneling regime.  Moreover, 
because the strength of the nonlinear self-interaction scales 
as $a_s/a_z$, by increasing the dimensions of the trap 
we reduce the effect of the nonlinearity thereby favoring 
quantum tunneling. 
\par 
In conclusion, we have shown that one can observe interesting 
macroscopic quantum-mechanical effects by studying the Bose-Einstein 
condensate in a double-well trap. We have accurately reproduced 
the formation of two Bose condensates observed by phase-contrast 
images at the MIT experiment [13]. 
Moreover, by using the Bogoliubov-de Gennes 
equations, we have analyzed the behavior of elementary 
excitations as a function of the strength of the double-well 
barrier. In particular, we have identified two 
collective excitations which have a different fate by increasing 
the strength of the barrier: an odd $q=0$ mode that 
asymptotically goes to zero and a even $q=1$ mode 
that remains constant also when the BEC 
separates in two condensates.  We hope that our calculations will
stimulate new, precise measurments of the excitation spectrum in this 
system in order to assess the validity of the theoretical framework
adopted in the theoretical analysis.  We have also considered 
the macroscopic quantum tunneling. Our calculations suggest that, 
in the tunneling region and with $^{23}$Na atoms, one sees only 
the macroscopic quantum self-trapping (MQST) 
of the condensate also when a small 
laser-sheet thickness is applied. We have shown that 
to get outside the MQST regime 
it is necessary to strongly reduce the scattering 
length or to increase the dimensions of the trap. Note 
that at nonzero temperature, BEC depletion and thermal fluctuations 
will slightly modify the parameters of the tunneling and will 
damp the coherent oscillations. 

\section*{Acknowledgements}
\par
This work has been supported by INFM under the Research Advanced Project (PRA) 
on "Bose-Einstein Condensation".

\begin{figure}
\centerline{\psfig{file=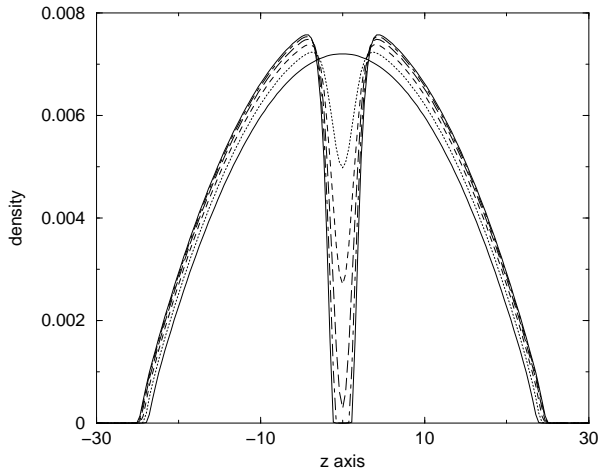,height=2.5in}}
\caption{Particle probability density 
in the ground state of $N=5\times 10^6$ 
$^{23}$Na atoms as a function of the $z$ axis at $r=0$ (symmetry plane). 
The curves correspond to increasing values of the strength $U_0$ 
of the barrier (from $0$ to $500$), in units of $\hbar \omega_z = 0.78$ peV. 
The laser power is given by the conversion formula 
$P=0.09 \times U_0$ mW. Lengths are in units of $a_z =4.63\; \mu$m.}  
\end{figure}

\begin{figure}
\centerline{\psfig{file=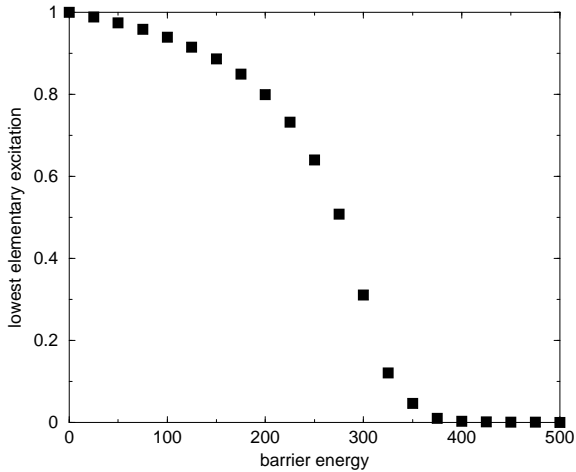,height=2.5in}}
\caption{Lowest elementary excitation $\hbar \omega_1^{(-)}$ with $q=0$ 
{\it vs} barrier energy $U_0$ for $N=5\times 10^6$ $^{23}$Na atoms. 
Units as in Fig. 1.}
\end{figure}

\begin{center}
\begin{tabular}{|ccccc|} \hline 
 $U_0$ & $E/N$ & $\mu$ & $\sqrt{<\rho^2>}$ & $\sqrt{<z^2>}$ \\ 
\hline 
$0$   & $211.142$ & $294.909$ & $0.989$ & $9.145$  \\
$100$ & $221.334$ & $303.910$ & $0.982$ & $9.635$  \\ 
$200$ & $227.467$ & $310.705$ & $0.986$ & $10.011$  \\
$300$ & $230.852$ & $315.297$ & $0.993$ & $10.268$ \\
$400$ & $232.855$ & $317.952$ & $0.997$ & $10.417$ \\
$500$ & $234.283$ & $319.763$ & $0.999$ & $10.518$ \\
\hline 
\end{tabular} 
\end{center} 

\vskip 0.3 truecm 
{\bf Table 1}. Ground state of $N=5\times 10^6$ $^{23}$Na atoms. 
Energies are in units of $\hbar \omega_z 
=0.78$ peV ($\omega_z=2\pi \times 19$ Hz). 
Lengths are in units of $a_z=4.63\;\mu$m. 
The laser power is given by the conversion formula 
$P=0.09 \times U_0$ mW. 

\vskip 0.5 truecm

\begin{center}
\begin{tabular}{|cccccc|} \hline
$U_0$ & $\hbar \omega_1^{(-)}$ 
& $\hbar \omega_2^{(+)}$ & $\hbar \omega_3^{(-)}$ 
& $\hbar \omega_4^{(+)}$ & $\hbar\omega_5^{(-)}$ \\ 
\hline
$0$   & $1.000$   & $1.580$ & $2.120$ & $2.645$ & $3.164$ \\
$100$ & $0.939$   & $1.596$ & $2.022$ & $2.640$ & $3.004$ \\
$200$ & $0.799$   & $1.624$ & $1.886$ & $2.711$ & $2.916$ \\
$300$ & $0.311$   & $1.643$ & $1.672$ & $2.724$ & $2.744$ \\
$400$ & $0.003$   & $1.655$ & $1.655$ & $2.730$ & $2.730$ \\ 
$500$ & $10^{-4}$ & $1.663$ & $1.663$ & $2.744$ & $2.744$ \\ 
\hline 
\end{tabular} 
\end{center} 

\vskip 0.3 truecm 
{\bf Table 2}. Lowest elementary excitations of the $q=0$ 
BdG spectrum for the ground state of $N=5\times 10^6$ $^{23}$Na atoms. 
$^{(-)}$ and $^{(+)}$ mean odd and even 
z-parity, respectively. Units as in Tab. 1. 

\vskip 0.5 truecm

\begin{center}
\begin{tabular}{|ccccc|} \hline
$U_0$ & $\hbar \omega_1^{(+)}$ 
& $\hbar \omega_2^{(-)}$ & $\hbar \omega_3^{(+)}$ 
& $\hbar \omega_4^{(-)}$ \\ 
\hline 
$0$   & $13.132$ & $13.165$ & $13.214$ & $13.278$ \\ 
$100$ & $13.132$ & $13.158$ & $13.218$ & $13.260$ \\ 
$200$ & $13.132$ & $13.145$ & $13.222$ & $13.236$ \\ 
$300$ & $13.132$ & $13.133$ & $13.225$ & $13.225$ \\ 
$400$ & $13.132$ & $13.132$ & $13.226$ & $13.226$ \\ 
$500$ & $13.132$ & $13.132$ & $13.227$ & $13.227$ \\ 
\hline 
\end{tabular} 
\end{center} 

\vskip 0.3 truecm 
{\bf Table 3}. Lowest elementary excitations of the $q=1$ 
BdG spectrum for the ground state of 
$N=5\times 10^6$ $^{23}$Na atoms. $^{(+)}$ and $^{(-)}$ mean even and odd 
z-parity, respectively. Units as in Tab. 1. 

\vskip 0.5 truecm

\begin{center}
\begin{tabular}{|cccc|} \hline 
$a_s/a_s^{Na}$ & $\Lambda$ & $\tau_0$ (sec) & $z_c$ \\ 
\hline 
$10^{-1}$  & $1108.337$ & $14.583$  & $0.060$ \\ 
$10^{-2}$  & $133.643$  & $13.887$  & $0.173$ \\ 
$10^{-3}$  & $1.390$    & $13.842$  & none    \\ 
$10^{-4}$  & $0.103$    & $10.253$  & none    \\ 
\hline 
\end{tabular} 
\end{center} 

\vskip 0.3 truecm 
{\bf Table 4}. Parameters of the MQT for different 
values of the scattering length $a_s$ with 
$a_s^{Na}=3$ nm and $\tau_0=2\pi \hbar/\Delta E^0$.  
Condensate with $N=5\times 10^3$ atoms. Barrier with 
$U_0=20$ and $\sigma= 1.5$ $\mu$m. Unit of $U_0$ as in Tab. 1. 

\end{document}